# "Plataforma Mobile de Aprendizado: Educação Adaptativa e Gamificada para Neuroatípicos"

Mirella E. B. Santana, Cauã O. Jordão, Victor B. Santos, Leonardo J. O. Ibiapina, Gabriel M. Silva, Marina R. Cavalcanti, Lucas R. C. Farias

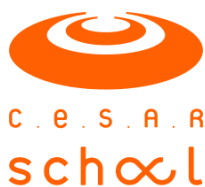

## 1. INTRODUÇÃO

A alfabetização é definida como o processo de aquisição de habilidades de leitura e escrita. Esse processo inicia-se pela exploração e pelo domínio da linguagem e se estende até a capacidade de codificar e decodificar os símbolos do alfabeto. Adicionalmente, é fundamental que o indivíduo seja capaz de compreender contextos e de aplicar o que lê e escreve. No entanto, no Brasil, a alfabetização ainda enfrenta muitos entraves, o que resulta em uma trajetória não linear entre crianças e jovens. Tal cenário é composto por múltiplos fatores, entre eles as desigualdades sociais e econômicas existentes no país, a escassez de recursos pedagógicos, os desafios estruturais do sistema de ensino e a metodologia atual, que não atende aos diversos perfis de aprendizagem.

Diante desse panorama, faz-se necessária a adoção de uma abordagem pedagógica multimodal, flexível e adaptativa, capaz de respeitar a heterogeneidade inerente às salas de aula. A curva de aprendizado é intrínseca a cada aluno, evidenciando que a construção do conhecimento ocorre de forma singular. Nesse sentido, Guerra (2000) reforça a perspectiva aqui apresentada, ao argumentar que crianças com dificuldades de aprendizagem não são incapazes, apenas precisam ser expostas a diferentes modalidades ou formas de apresentação do conhecimento.

Com o intuito de atender à demanda por estratégias inovadoras no ensino, observa-se o crescimento de pesquisas e estudos na área, que envolvem principalmente o uso de novas tecnologias, como aplicativos interativos, jogos educativos e sistemas baseados em Inteligência Artificial (IA). Para a educação, a aliança com a tecnologia apresenta grande potencial para personalizar e otimizar o ensino-aprendizagem, principalmente por permitir a integração e a manipulação de diversos elementos sensoriais e cognitivos, incluindo o uso de sons, cores e imagens, o estímulo a recursos táteis e o incentivo a atividades de coordenação motora fina. Essa variedade de possibilidades facilita a exploração de múltiplos canais de informação pelo estudante, o que é essencial para o avanço da educação inclusiva e mais equitativa.

Considerando o exposto, o primeiro pilar argumentativo desta pesquisa baseia-se no reconhecimento de que o ensino tradicional não consegue atender a todos os indivíduos de forma igualitária. A alfabetização para crianças neuroatípicas costuma apresentar padrões distintos daqueles observados em perfis neurotípicos, pois existem variações neurobiológicas que impactam no processamento de informações, na capacidade de memorização e na atenção dada ao conteúdo, habilidades cognitivas essenciais para este período de desenvolvimento. Sob esta ótica, submeter uma criança neuroatípica ao formato habitual de educação não é o modelo ideal. Consequentemente, a ausência de um ensino personalizado contribui para uma exclusão sistêmica, uma vez que não compromete apenas a permanência e o desempenho nas salas de aula, mas também resulta na restrição de oportunidades sociais para esses alunos, culminando, em muitos casos, na transição para a vida adulta com defasagem no aprendizado ou com analfabetismo completo.

Atualmente, encontra-se na literatura uma vasta quantidade de pesquisas sobre o uso de jogos sendo colocados como apoio para a alfabetização e o letramento em sala de aula, principalmente nas séries iniciais da educação básica. Porém, ao pesquisarmos por artigos que tratem especificamente da alfabetização para o público neuroatípico, esse número cai significativamente, o que evidencia a escassez de investigações científicas e a carência de soluções tecnológicas voltadas a essas singularidades. Essa lacuna demonstra que, embora a gamificação já seja aceita e colocada em prática, sua aplicação ainda carece de uma especialização metodológica que contemple o processamento cognitivo divergente. Nota-se que a maioria das aplicações disponíveis foca em atividades lúdicas isoladas, carecendo de um sistema integrado, que permita a portabilidade de intervenções específicas, como o método AFA, para o ambiente móvel de forma dinâmica e acessível. É precisamente nesta intersecção que este estudo se insere.

A sustentação da hipótese principal da nossa tese provém da lacuna identificada e exposta anteriormente: a rigidez dos métodos tradicionais não se adapta às singularidades neurocognitivas de crianças neuroatípicas. O segundo viés apresentado, o potencial das novas tecnologias voltadas ao ensino, sustenta a tese ao servir como uma alternativa para a questão, visto que estas oferecem um mecanismo robusto para a criação de novas metodologias. Assim, estabelecemos como Hipótese de Pesquisa (H) que a implementação de uma aplicação similar a um jogo, com recursos multissensoriais e trilhas de atividades customizadas, durante a alfabetização de um estudante neurodivergente, será capaz de obter resultados superiores quando comparada a abordagens pedagógicas convencionais. Portanto, a Questão de Pesquisa (Q) que este estudo se propõe a responder pode ser definida como: "Em que medida a utilização de uma plataforma mobile gamificada e adaptativa demonstra maior eficácia na aquisição de competências de alfabetização e no nível de engajamento de crianças neuroatípicas, em comparação com os métodos tradicionais de ensino?"

A fim de validar a hipótese de pesquisa, o objetivo geral deste plano de atividades é desenvolver um aplicativo mobile intuitivo e acessível (denominado Wing), que utilize o aprendizado adaptativo e elementos de gamificação para aprimorar o processo educacional de crianças e jovens neuroatípicos. Como objetivos específicos, temos: criar uma interface acessível, garantindo usabilidade para diferentes perfis de alunos; digitalizar e integrar o método AFA, tornando-o interativo e dinâmico no ambiente digital; incorporar artefatos multisensoriais, como cores, sons e estímulos visuais, para aumentar o engajamento dos usuários; desenvolver trilhas de conhecimento personalizadas, adaptando o conteúdo ao progresso e às necessidades individuais de cada usuário; criar um sistema de gamificação baseado em desafios e recompensas que incentive a progressão no aprendizado; e conduzir testes com especialistas e usuários, refinando a experiência do aplicativo a partir dos feedbacks obtidos.

A principal contribuição desta pesquisa reside em enfrentar a carência de recursos didáticos multimodais e personalizados, que auxiliem na trajetória de alunos com necessidades específicas. É válido ressaltar que a solução aqui apresentada não se propõe a substituir integralmente a alfabetização tradicional ou o papel do educador, mas a atuar como uma ferramenta complementar de apoio, visando amenizar esse hiato educacional e potencializar os resultados pedagógicos. Ademais, o projeto tem como intenção digitalizar o método AFA, conferindo-lhe um caráter escalável e de amplo alcance, superando as barreiras geográficas. Espera-se também que a natureza gamificada e adaptativa da plataforma resulte em maior engajamento e autonomia do aluno durante o processo de aprendizagem, uma vez que o uso de gamificação demonstra eficácia ao capitalizar o interesse natural dos jovens por jogos, favorecendo a aquisição de habilidades de forma prazerosa e imersiva.

O presente relatório está organizado em cinco seções. Iniciamos com uma Introdução, na qual contextualizamos a temática e definimos os objetivos. Em seguida, a Revisão da Literatura fundamenta a tese, discutindo os estudos correlatos. O procedimento adotado é detalhado na seção de Materiais e Métodos. Posteriormente, dedicamos uma seção à Análise e Discussão dos Resultados obtidos e, finalmente, a Conclusão sintetiza os principais achados e aponta as perspectivas futuras deste projeto.

## 2. Fundamentação Teórica

### 2.1. Neurodiversidade e Acessibilidade Digital

A neurodiversidade refere-se à variação natural no funcionamento cognitivo humano, incluindo condições como Transtorno do Espectro Autista (TEA), Transtorno de Déficit de Atenção e Hiperatividade (TDAH), dislexia e discalculia. Indivíduos neurodivergentes possuem diferentes formas de processar informações, o que exige adaptações específicas nos ambientes digitais (PASCHOALOTTO; JUSTO, 2024). Estudos indicam que 15% a 20% da população mundial apresenta alguma forma de neurodivergência, demonstrando a relevância numérica desse público (PASCHOALOTTO; JUSTO, 2024).

No contexto digital, pessoas neurodivergentes enfrentam barreiras significativas como: dificuldade em compreender conteúdos não apresentados de forma clara, distração com excesso de elementos visuais, navegação complexa e falta de recursos de acessibilidade adequados (PASCHOALOTTO; JUSTO, 2024). A Lei Brasileira de Inclusão (LBI/2015) torna obrigatória a acessibilidade digital, estabelecendo que conteúdos online devem ser perceptíveis, operáveis, compreensíveis e robustos para todos os usuários.

## 2.2. Diretrizes para Design Inclusivo

Para atender às necessidades específicas do público neurodivergente, este projeto fundamenta-se em dois conjuntos principais de diretrizes: o GAIA (Guia de Acessibilidade de Interfaces para Autismo) e as recomendações do projeto .horeel.

O GAIA oferece 28 recomendações para desenvolvimento de interfaces acessíveis, abordando aspectos como vocabulário visual, customização, engajamento e representações redundantes de informação (GAIA, 2018; SERATI; GIBERTONI, 2022). Entre suas orientações estão: uso de linguagem simples e objetiva, previsibilidade na navegação, feedback claro das ações e eliminação de elementos distrativos.

O guia .horeel (2020) complementa essas recomendações com foco específico em TDAH, dislexia, discalculia e disortografia, sugerindo práticas como: uso de tipografias de fácil leitura (como OpenDyslexic ou Dyslexie), contraste adequado de cores, layout limpo com hierarquia visual clara, e minimização de animações e movimento (PASCHOALOTTO; JUSTO, 2024).

A integração dessas diretrizes resultou no Guia de Boas Práticas WING (Figura 1), documento que orientou todas as decisões de design do aplicativo, garantindo uma abordagem inclusiva desde a concepção inicial.

## 2.3. Experiência do Usuário (UX) Aplicada ao Desenvolvimento Inclusivo

A Experiência do Usuário (UX) desempenha papel fundamental no desenvolvimento de produtos para públicos específicos. Segundo o modelo da "Colmeia da Experiência do Usuário" de Morville (2004), acessibilidade representa uma das sete facetas essenciais da experiência, ao lado de utilidade, usabilidade, desejabilidade, encontrável, credibilidade e valor (SERATI; GIBERTONI, 2022).

No contexto de desenvolvimento para neurodivergentes, a UX requer abordagens especializadas. Serati e Gibertoni (2022) demonstram a eficácia da combinação entre avaliações por inspeção específicas (utilizando diretrizes como GAIA) e heurísticas gerais de usabilidade (como as 10 Heurísticas de Nielsen) para identificar problemas e orientar melhorias no design.

Esta abordagem mista permite avaliar tanto aspectos específicos da acessibilidade para o público-alvo quanto características gerais de usabilidade, garantindo que o produto final seja não apenas acessível, mas também intuitivo, engajador e eficaz em seu propósito pedagógico. A aplicação desses princípios no desenvolvimento do Wing assegura que o aplicativo atenda às necessidades cognitivas específicas das crianças neuroatípicas enquanto mantém padrões elevados de qualidade na experiência do usuário.

## 3. Materiais e Métodos

O estudo é de natureza aplicada, pois se propõe a resolver um problema real por meio da criação de um produto e adota uma abordagem mista, combinando métodos qualitativos e quantitativos. Possui caráter qualitativo ao interpretar dados não numéricos, como observações dos educadores sobre o comportamento do aluno durante o uso do aplicativo, feedbacks de especialistas da área e aderência dos alunos às atividades. Simultaneamente, o quantitativo é evidenciado pela coleta de informações como taxa de engajamento, número de acertos e progresso de fases, que são registradas à medida que os usuários avançam nas trilhas de aprendizagem. Para fundamentar teoricamente a pesquisa, foram realizadas revisões bibliográficas, análises de concorrentes e entrevistas com a psicopedagoga Cláudia Rêgo. A produção de software é o principal artefato, e as coletas de dados numéricos devem ser realizadas na fase de testes da aplicação. O universo da pesquisa abrange o ecossistema educacional de crianças e jovens neuroatípicos, de educadores e de familiares que atuam no processo de alfabetização.

A condução da análise dos dados será dada através de uma estratégia de triangulação de métodos, integrando as perspectivas qualitativa e quantitativa para uma melhor compreensão do objeto estudado. Para os dados quantitativos, obtidos através da experiência dos alunos (taxa de acerto de questões, progresso nas fases, tempo utilizado em tarefas e métricas de engajamento com a aplicação), serão empregadas estatísticas descritivas e análise correlacional, permitindo mensurar de forma objetiva o desempenho e a interação dos usuários com o aplicativo. Paralelamente, os dados qualitativos - provenientes de observações dos educadores, feedbacks de especialistas e relatos das famílias dos alunos - serão submetidos à análise de conteúdo temática, buscando identificar padrões, dificuldades recorrentes e percepções sobre a usabilidade e a eficácia pedagógica da ferramenta. A escolha por esta abordagem mista de análise justifica-se pela complexidade do objeto de pesquisa, que envolve tanto aspectos mensuráveis de desempenho quanto dimensões subjetivas de experiência, comportamento e aprendizagem. Dessa forma, a triangulação assegura que os resultados não se restrinjam aos números, mas sejam interpretados e contextualizados pelo viés humano e especializado.

A proposta central desta pesquisa é o desenvolvimento e a validação de um aplicativo digital que transponha de forma eficaz o método AFA para um ambiente interativo, promovendo a alfabetização de forma mais engajadora e adaptativa, através da gamificação e da personalização para os diversos perfis de crianças e jovens neuroatípicos. Para validar essa proposição, é fundamental demonstrar não apenas a funcionalidade técnica do software, mas, sobretudo, o seu impacto pedagógico e comportamental. É nesse ponto que a escolha do método de análise de dados mostra-se diretamente alinhada ao objetivo. A análise quantitativa fornecerá evidências concretas sobre a progressão na aprendizagem e o nível de engajamento dos alunos, indicadores objetivos da eficiência da ferramenta. Já a análise qualitativa será crucial para capturar as nuances do processo: como a criança se relaciona afetivamente com a atividade, quais desafios persistem e como a ferramenta pode estar mediando novos comportamentos de aprendizagem. Portanto, a análise mista é o que permitirá responder à pergunta de pesquisa de forma completa, conectando a eficácia mensurável do produto (o como e quanto) à sua significação no contexto real de uso (o porquê e de que forma), validando assim a proposta de valor da solução desenvolvida.

A forma de coleta de dados será estruturada por meio de testes de usabilidade e de funcionalidade com o público-alvo, utilizando uma abordagem de estudo de caso, com duração estimada de quatro semanas. A amostra utilizada será composta por 20 alunos já integrados ao método AFA, para os quais os responsáveis autorizam a participação no programa. O procedimento consistirá na execução das primeiras trilhas disponibilizadas pelo nosso time de desenvolvimento aos estudantes. Após a finalização deste mês de testes, os feedbacks serão coletados, assim como mapeados os indicadores que definimos como critérios quantitativos. Adicionalmente, os dados coletados serão avaliados, de forma técnica e pedagógica, pela especialista Cláudia Rêgo, para validar a eficácia da transposição do método para o ambiente digital e a resposta comportamental dos alunos durante o uso da ferramenta.

Para a construção do instrumento de pesquisa, utilizamos um fluxo de Design Centrado no Usuário. Primeiramente, foram elencadas as diretrizes a serem seguidas para a elaboração de um design inclusivo. Em seguida, foram feitas entrevistas para compreensão e análise dos requisitos pedagógicos baseados no método AFA, o que garantiu que a lógica de alfabetização fosse preservada na transição para o meio digital. O protótipo de alta fidelidade foi estruturado com o auxílio da plataforma Figma, o que possibilitou o início da codificação e o surgimento das primeiras telas. A implementação das trilhas de aprendizagem foi estruturada sob uma lógica de aprendizado adaptativo, o que faz com que o sistema ajuste o nível de dificuldade conforme a interação do aluno, atuando como o meio pelo qual os dados de desempenho e engajamento serão extraídos durante a fase de experimentação.

Para validar os resultados obtidos após o uso dos métodos propostos, os dados quantitativos de desempenho extraídos da plataforma serão comparados às análises clínicas da especialista psicopedagoga e aos relatos dos pais, por meio da triangulação de dados. Essa abordagem permite verificar se a evolução registrada digitalmente se traduz em ganho real de autonomia e de competência cognitiva. Por fim, a eficácia da solução será aferida pela correlação entre o aumento do engajamento (tempo de uso e recorrência) e a taxa de sucesso nas atividades propostas, fornecendo uma base sólida para confirmar ou refutar a hipótese central.

## 4. Análise e discussão de resultados

### 4.1. Fase Inicial — Pesquisa e Levantamento de Requisitos

Na etapa inicial do projeto, realizamos pesquisas exploratórias e teóricas sobre as temáticas centrais da nossa proposta, com o objetivo de aprofundar o entendimento e estabelecer uma base conceitual sólida para o desenvolvimento do aplicativo. Buscamos compreender os princípios da educação inclusiva, do design adaptativo e das tecnologias assistivas, a fim de garantir que a solução atenda às necessidades do público-alvo. A partir desse estudo, elaboramos um guia de boas práticas de design, fundamentado nas diretrizes HORCEL e GAIA, voltado a assegurar uma experiência do usuário acessível, intuitiva e inclusiva. Esse guia reúne orientações detalhadas sobre o uso de fontes, a paleta de cores, a escolha de imagens e de elementos visuais, servindo como referência para manter a consistência e a acessibilidade em toda a interface do aplicativo.

**Figura 1 –** Guia WING de boas práticas no desenvolvimento de aplicativos inclusivos

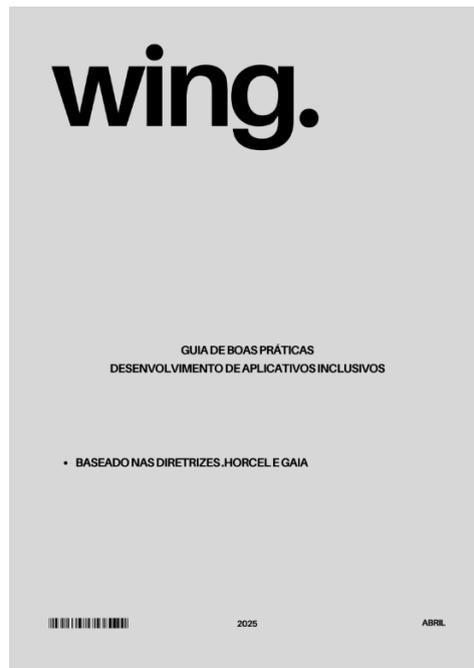

**Fonte:** Autores, 2025.

Paralelamente, realizamos uma análise de aplicativos e plataformas voltadas à alfabetização, com o intuito de mapear o cenário atual e identificar possíveis concorrentes da nossa solução. Essa etapa permitiu compreender as abordagens existentes e identificar oportunidades de inovação. Além disso, foi necessário entendermos o método AFA (Alfabetização Adaptada), com o objetivo de compreender suas etapas, recursos e fundamentos pedagógicos, a fim de traduzir a metodologia do formato físico para o ambiente digital. O apoio da psicopedagoga Cláudia Rêgo, criadora do método, foi essencial nesse processo, contribuindo diretamente para o alinhamento entre as propostas tecnológicas e os princípios educacionais do AFA. Para isso, realizamos reuniões dedicadas ao aprendizado do método, nas quais pudemos esclarecer dúvidas e compreender sua aplicabilidade.

**Figura 2 –** Reunião para aprendizagem do método AFA

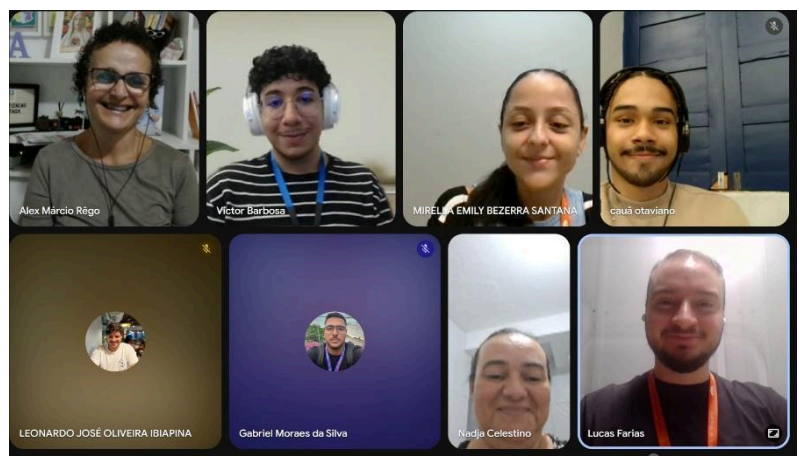

Após essas etapas, foi possível levantar e documentar os requisitos da solução, que foram organizados e acompanhados na ferramenta Jira. A partir disso, iniciamos o desenho da arquitetura da aplicação e o processo de prototipação das telas iniciais, via Figma.

## 4.2. Prototipação — Design da Interface e Fluxo de Navegação

Com base nos requisitos levantados, iniciou-se a criação do protótipo de alta fidelidade no Figma, que representa as principais funcionalidades do aplicativo. Essa etapa permitiu validar a organização das telas, o fluxo de navegação e a disposição dos elementos visuais antes de sua implementação.

**Figuras 3, 4, 5, 6, 7 e 8** – Telas do protótipo no Figma

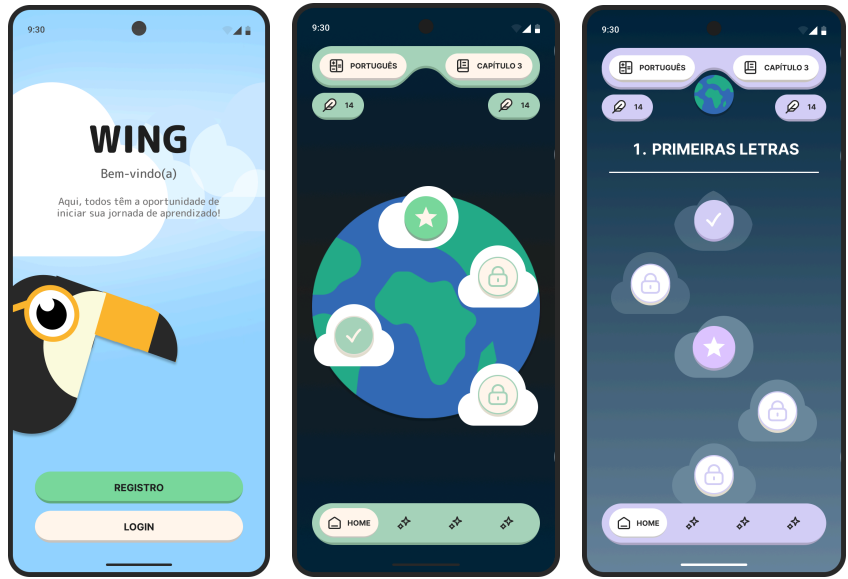

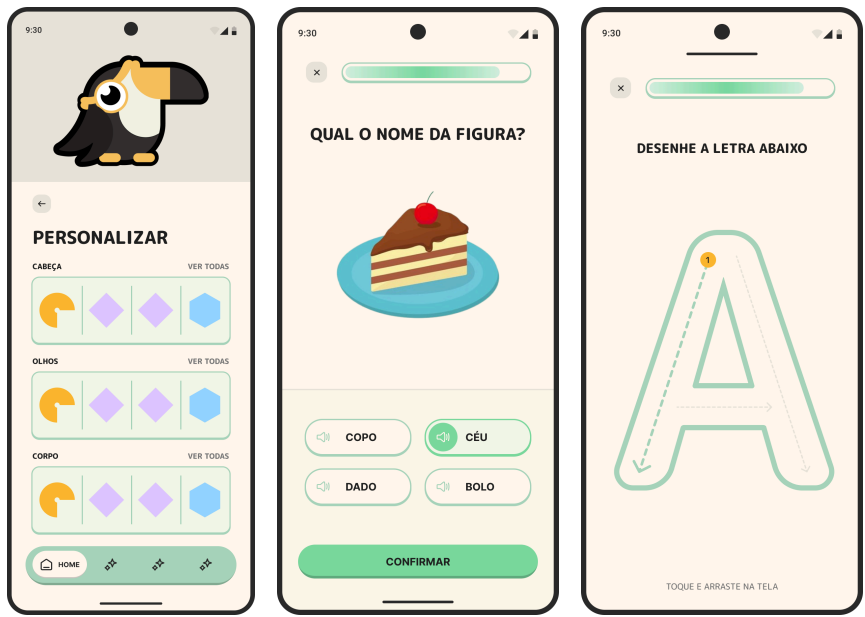

**Fonte:** Autores, 2025.

### 4.3. Divulgação Social — Participação em Eventos da Comunidade Científica

Durante os dias 15, 16 e 17 de outubro de 2025, participamos do **REC'n'Play**, festival de inovação, tecnologia e cultura no Recife, no espaço do **FORJA**, projeto de desenvolvimento de jogos realizado por estudantes da CESAR School. A participação no evento proporcionou uma rica interação com o público, permitindo receber diversos feedbacks, compartilhar experiências e explorar possíveis parcerias futuras. Esse ambiente foi essencial para validar nossa proposta, fortalecer a visibilidade do Wing e conectar o projeto com a comunidade acadêmica e tecnológica.

**Figuras 9 e 10 –** Participação no REC'n'Play 2025

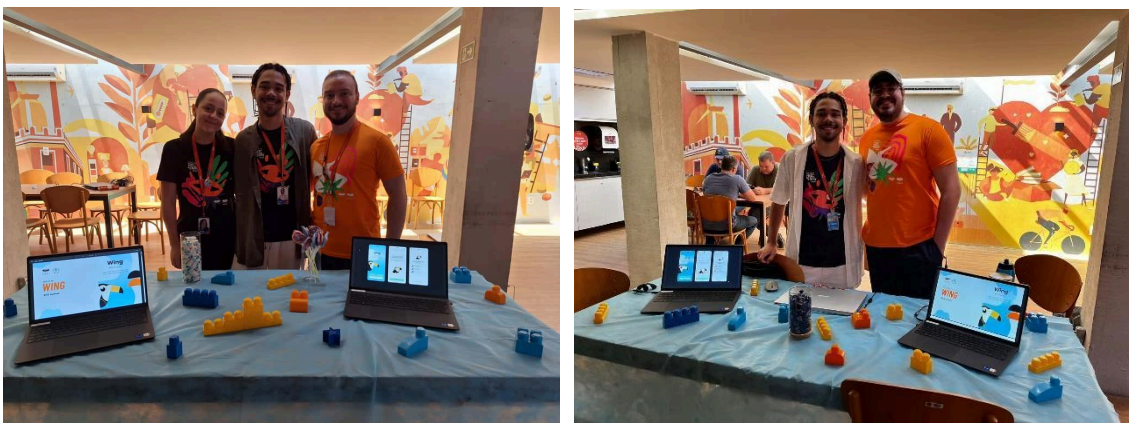

**Fonte:** Autores, 2025.

Em 13 de dezembro de 2025, o Wing esteve presente na Mostra Tech, evento semestral onde alunos de graduação apresentam projetos práticos desenvolvidos para desafios reais, unindo tecnologia, design e inovação. A participação neste evento marcou a finalização do projeto de Iniciação Científica 2025, fundamentando os próximos passos para a implementação e testes de campo previstos para a próxima fase do estudo.

**Figura 11** – Participação na Mostra Tech 2025

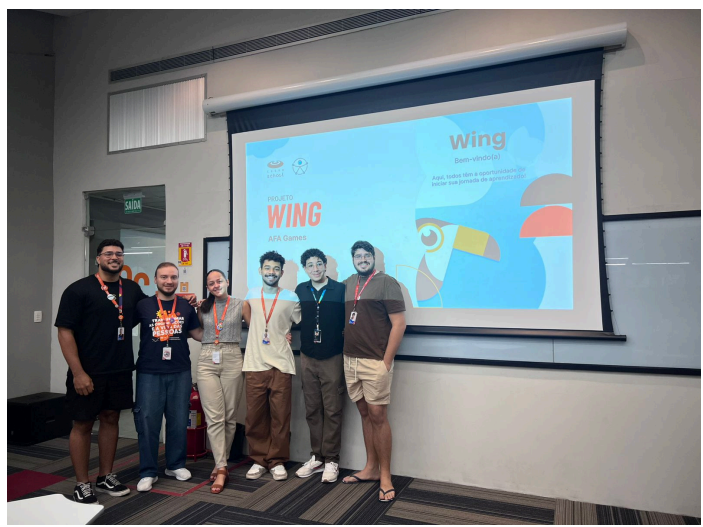

**Fonte:** Autores, 2025.

## 5. Conclusão

Esta pesquisa foi motivada pela necessidade de atender às particularidades de ensino para crianças neuroatípicas no processo de alfabetização. O problema central identificado foi a carência de ferramentas digitais que integrem métodos pedagógicos consolidados a uma interface verdadeiramente inclusiva, gamificada e adaptada. Diante disso, o objetivo deste estudo foi desenvolver o Wing, um aplicativo mobile fundamentado no método AFA, projetado para oferecer uma jornada de aprendizagem personalizada, capaz de respeitar o ritmo cognitivo de cada aluno.

Utilizando uma abordagem de pesquisa aplicada, prototipação e metodologias ágeis, o primeiro resultado concreto foi o nosso MVP (Mínimo Produto Viável). Esta entrega permitiu a validação prática das premissas de interface e acessibilidade durante a Mostra Tech 2025, onde o projeto foi submetido à avaliação da banca. Somado a isso, por meio da revisão bibliográfica aliada à consultoria técnica com a especialista Claudia Rêgo, o estudo alcançou a transposição do método AFA para o ambiente digital. Este resultado evidencia a viabilidade de converter uma intervenção tradicional, sem perder o rigor pedagógico necessário.

Ao revisitar o objetivo de pesquisa, constata-se o cumprimento da meta de desenvolvimento do aplicativo através da entrega do MVP (Minimum Viable Product), o qual materializa as premissas de design inclusivo e acessibilidade elencadas inicialmente. É importante ressaltar, contudo, que a elaboração das trilhas de aprendizagem ainda se encontra em fase de execução. A conclusão desta etapa é o requisito final para o início da fase de experimentação e testes de campo com os usuários, momento em que a eficácia pedagógica da ferramenta será formalmente mensurada.

Pode-se concluir que o Wing vai além da concepção de um simples aplicativo, pois atua como uma ferramenta capaz de gerar impacto na realidade social e pedagógica de seus usuários. Ao digitalizar o método do AFA, o Wing contribui para a democratização do acesso a um ensino técnico e especializado, frequentemente limitado por barreiras financeiras ou geográficas enfrentadas por muitas famílias. Destaca-se também a promoção da autonomia do aluno neuroatípico, que deixa de ocupar uma posição passiva em seus estudos para assumir o protagonismo de sua própria trajetória de aprendizagem. Dessa forma, a pesquisa contribui diretamente para o fortalecimento da autoestima de crianças que, anteriormente, encontravam-se à margem dos processos tradicionais de alfabetização.

Como sugestão para trabalhos futuros, recomenda-se a realização de análises dos resultados obtidos, utilizando uma amostra ampliada de usuários, com o objetivo de mensurar os ganhos de aprendizado a longo prazo de forma comparativa. Ademais, outra oportunidade que observamos é a exploração de algoritmos de IA mais avançados para o refinamento do aprendizado adaptativo. Por fim, sugerimos também a expansão do aplicativo para contemplar outras áreas do conhecimento além da alfabetização inicial.